# Evolution of structure, magnetism, and electronic/thermal-transports of Ti(Cr)-substituted Fe$_2$CrV all-d-metal Heusler ferromagnets

Yiting Feng[1,2], Shen Zhang[1], Qingqi Zeng[1], Meng Lyu[1], Junyan Liu[1], Jinying Yang[1,2], Yibo Wang[1,2], Qiusa Ren[1,3], Yang Liu[1,2], Binbin Wang[1], Hongxiang Wei[1], Enke Liu[1*]

[1] Beijing National Laboratory for Condensed Matter Physics, Institute of Physics, Chinese Academy of Sciences, Beijing 100190, China;
[2] School of Physical Sciences, University of Chinese Academy of Sciences, Beijing 100049, China;
[3] University of Science and Technology Beijing, Beijing 100083, China.

* Email: ekliu@iphy.ac.cn

**Abstract**

All-d-metal full-Heusler alloys possess superior mechanical properties and high spin polarization, which would play an important role in spintronic applications. Despite this, their electrical and thermal transport properties have not been comprehensively investigated till now. In this work, we present an analysis on the evolution of structural, magnetic and transport properties of Cr- and Ti-substituted Fe$_2$CrV all-d-metal Heusler alloys by combining theoretical calculations and experiments. Both series of alloys crystallize in Hg$_2$CuTi-type structure. With increasing Ti doping, the calculated total magnetic moments of Fe$_{50}$Cr$_{25}$V$_{25-x}$Ti$_x$ decrease linearly. The experimental saturation magnetization is highly consistent with theoretical calculations and Slater-Pauling rule when x < 4, indicating the highly ordered atomic occupation. The magnetization and Curie temperature can be significantly tuned by altering spin polarizations and exchange interactions. The introduction of the foreign atom, Ti, results in a linear increase in residual resistivity, while electron-phonon scattering keeps relatively constant. The maximum values for electrical and thermal transport properties are observed in the stoichiometric Fe$_2$CrV composition.

## 1. Introduction

The Heusler compounds, originally investigated by the German chemist Friedrich Heusler[1], have been extensively studied for several decades due to their diverse and significant physical properties, including shape memory effect[2-5], magnetoresistance[6-8], magnetocaloric effect[9, 10], topological insulators[11-13] and superconductivity[14-16].

These properties render the Heusler alloys as promising candidates for technological applications such as solid-state refrigerants[17], spintronics[18] and quantum devices[19]. The full-Heusler ternary compounds follow the composition $X_2YZ$, where X and Y represent transition metal elements and Z represents main-group elements. They offer the advantages of easy synthesis, structural stability and high Curie temperature ($T_C$).

The concept of all-d-metal full-Heusler alloys (X, Y represent transition metal elements with more valence electrons and Z transition metal elements with less valence electrons) with d−d covalent hybridization was firstly proposed by Wei et al. in 2015[20]. Low-valence Ti was introduced into the binary NiMn alloy, creating the $Ni_{50}Mn_{50-y}Ti_y$ and $Mn_{50}Ni_{50-y}Ti_y$ systems. Building on this, the ferromagnetic martensitic transformations were observed in $Ni_{50-x}Co_xMn_{50-y}Ti_y$ systems by doping with Co atoms. In addition, a large magnetic field-induced strain and volume change were achieved in $Mn_{50}Ni_{40-x}Co_xTi_{10}$ alloys[21]. These phase-transition-based multifunctional properties and superior mechanical properties make the novel class of material systems highly interesting for multi-driven applications. The evolution of the electronic structure during the martensitic transformations were investigated by electronic transport measurement[22]. Subsequently, giant barocaloric effect, magnetocaloric effect and elastocaloric effect were reported in various all-d-metal Heusler alloys, which exhibits the potential use for clean solid-state refrigeration technologies[2, 3, 23-25]. It is evident that most studies have focused on the magnetic phase transitions, with only a few studies on transport properties. The ab-initio electronic structure calculations revealed that the novel class of magnetic Heusler compounds usually half-metallicity and high spin polarization at the $E_F$, making them highly promising for spintronic applications[26]. This unique behavior attributed to the opening of a gap in one of the spin bands within the material[27]. It is important to expand the research on all-d-metal Heusler alloys and search for new physical behaviors and properties, including atomic occupation and electric-thermal transport.

The crystal structure and magnetic properties of new all-d-metal Heusler alloys $Fe_{50-x}Cr_{25+x}V_{25}$ (x=0–25) was investigated systematically in our previous work[28]. There existed an abnormal decrease of lattice constant, which is induced by the atomic configuration. Moreover, the $T_C$ can be remarkably tuned by substituting. Although these behaviors provide indications for understanding the all-d-metal Heusler alloys, the deeper properties, such as band structures and transport properties, are still unknown. There is limited research on the electric and thermal transport properties of all-d-metal Heusler alloys with d−d covalent hybridization. Hence, an investigation on the transport properties would provide further understanding and deepen our comprehension of this novel class of Heusler alloys.

In the present work, we first studied the evolution of the magnetization, Curie temperature, electronic structure of Ti-substituted $Fe_2CrV$ alloys. By using theoretical calculations and experiments, we then investigated the transport properties of both Ti- and Cr-substituted $Fe_2CrV$ alloys to further reveal the electronic characteristics. Considering the previous study of Cr-substituted $Cr_{75-x}Fe_xV_{25}$, more attentions were paid to Ti-substituted $Fe_{50}Cr_{25}V_{25-x}Ti_x$ alloys, with electronic and thermal transport of former system being studied in the present work.

## 2. Experiments

Two series of polycrystalline alloys, namely $Cr_{75-x}Fe_xV_{25}$ (x=25-50) and $Fe_{50}Cr_{25}V_{25-x}Ti_x$ (x=0-8), were synthesized by arc melting the high purity (99.99% or higher) elementary metals of Fe, Cr, V, Ti in argon atmosphere. The ingots were repeatedly melted at least four times to ensure homogeneity. Subsequently, they were sealed in evacuated quartz tubes and annealed at 773 K for 5 days to achieve homogenization and atomic ordering, and then quenched into ice water. The melt-spun ribbons of $Fe_{50}Cr_{25}V_{25-x}Ti_x$ (x=0-8) were prepared by a single wheel technique with the substrate velocity of 25 m/s, under the protection of argon atmosphere. RT X-ray diffraction (XRD) analyses were recorded on the flat surface of all sample plates using Cu-Kα radiation. The XRD patterns of ribbons were obtained by pasting them on a 10 mm × 10 mm glass sheet. The exact elemental compositions of the samples were detected by X-ray energy dispersive spectroscopy (EDS). $T_C$ was determined by Thermogravimetry (TG) Analysis of Simultaneous Thermal Analyzer (STA). As the ferromagnetic material is heated in the presence of a magnetic field, the magnetic properties disappear and the reduced attraction for the magnet results in a sharp apparent weight change at the $T_C$. Magnetic properties were measured by superconducting quantum interference device (SQUID) magnetometer at 5 K. Electrical transport properties were measured by physical property measurement system (PPMS). Thermoelectric transport measurements were performed in a PPMS cryostat. The temperature gradient was generated by a 1000 Ω resistive heat source and a copper heat sink at both ends of the sample. The temperature difference $\Delta T$ was set to 2-3% of the base temperature and measured by an Au-Fe thermocouple. $\Delta T$ was applied along the longitudinal axis, while the magnetic field was applied perpendicular to the sample. To eliminate the effect of contact misalignment, the data collected were subjected to field-symmetrization and antisymmetrization. The theoretical molecular magnetic moments and electronic structures were calculated by the Korringa-Kohn-Rostoker coherent potential approximation based on the local density approximation (KKR-CPA-LDA).

## 3. Results and discussion

### 3.1. Crystal Structure

The Cr- and Ti- substituted $Fe_2CrV$ alloys crystallize in XA-type ($Hg_2CuTi$-type with space group $F\bar{4}3m$, No. 216) structure[28] with the atomic configuration shown in Fig. 1(a). The unit cell is composed of four interpenetrating face centered cubic (fcc) sublattices A, B, C and D arranged along the diagonal line. The Wyckoff coordinates are A (0, 0, 0), B (0.25, 0.25, 0.25), C (0.5, 0.5, 0.5) and D (0.75, 0.75, 0.75), respectively. According to the atomic occupation rule, also known as Burth's rule[29], the paths of substituting Fe with Cr as well as V with Ti are shown in Fig. 1(b). The exact elemental compositions of $Fe_{50}Cr_{25}V_{25-x}Ti_x$ (x=0-8) annealed samples (denoted as Tix) from EDS are listed in Table 1. The composition is coincided with the nominal formula within the permitted margin of error. Figure 1(c) shows the RT XRD patterns of annealed bulk and ribbon samples. The starting alloy $Fe_2CrV$ crystallizes in cubic B2 phase, which is in good agreement with our previous work[1]. Upon substitution of Ti for V, extra peaks start to appear except for the main peaks of the bcc structure when x increases to 6. The extra peaks can be indexed to the FeTi cubic phase (space group $Pm\bar{3}m$, No. 221). The secondary phase, which is not eliminated by annealing, is inhibited by melt spinning. The content of secondary phase in ribbons is smaller than in annealed bulks.

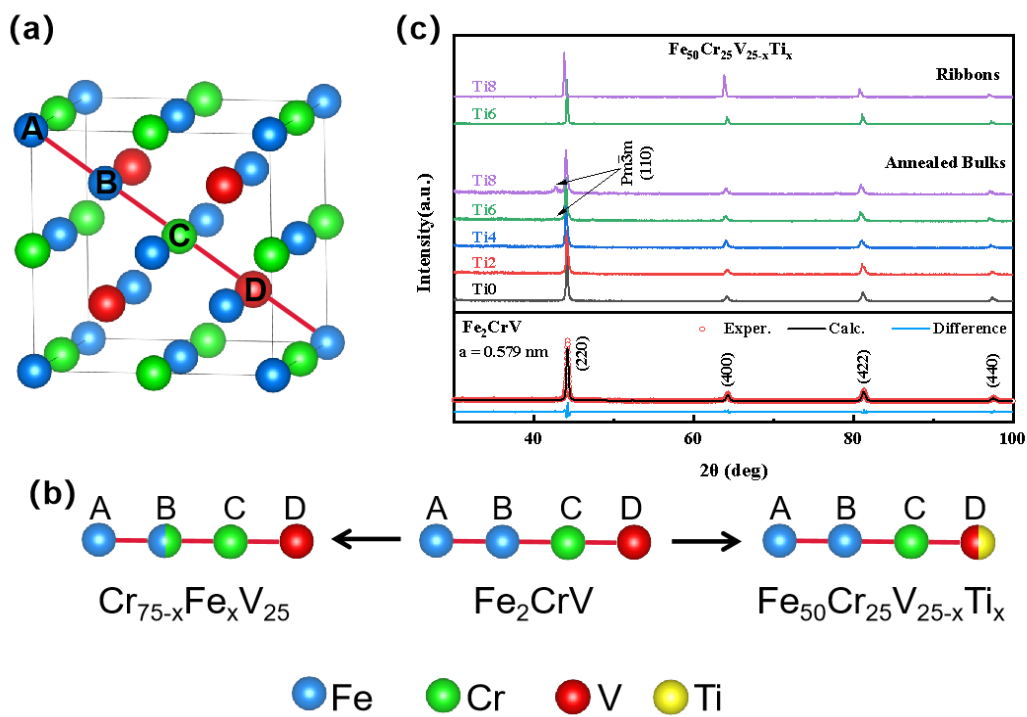

**Fig. 1.** (**a**) Models of the $Hg_2CuTi$-type (space group $F\bar{4}3m$) structure, in which A, B, C and D represent four interpenetrating fcc sublattices on the diagonal line. (**b**) Paths of substituting Fe with Cr as well as V with Ti. (**c**) (upper panel) RT XRD patterns of $Fe_{50}Cr_{25}V_{25-x}Ti_x$ (x=0-8, denoted as Tix) annealed bulks and ribbon samples. (Lower panel) The experimental, calculated and comparative XRD patterns of $Fe_2CrV$ sample.

| Nominal x | Actual concentration |
|---|---|
| 0 | $Fe_{48.49}Cr_{26.53}V_{24.97}$ |
| 2 | $Fe_{51.08}Cr_{23.95}V_{22.57}Ti_{2.40}$ |
| 4 | $Fe_{50.20}Cr_{24.65}V_{20.85}Ti_{4.30}$ |
| 6 | $Fe_{50.43}Cr_{24.22}V_{18.82}Ti_{6.53}$ |
| 8 | $Fe_{50.04}Cr_{24.74}V_{17.03}Ti_{8.18}$ |

**Table 1** The exact elemental compositions of $Fe_{50}Cr_{25}V_{25-x}Ti_x$ (x=0-8) annealed samples from EDS.

### 3.2. Magnetic Properties

Figure 2(a) shows the magnetization curves $M(H)$ at 5 K along substitution paths for $Fe_{50}Cr_{25}V_{25-x}Ti_x$ alloys, with corresponding experimental and calculated molecular magnetic moment ($M_s$) in Fig. 2(b). $M(H)$ curves exhibit soft ferromagnetic behavior, with saturation fields of approximate 1 kOe. With increasing Ti doping, the calculated $M_s$ of the system decreases monotonically. This is attributed to the introduction of Ti, the magnetic moments of Cr atoms align antiparallel with those of Fe atoms, adding a local FIM structure to the native FM matrix. The measured $M_s$ are highly consistent with the theoretical results and the Slater-Pauling rule when x < 4, indicating the highly ordered atomic occupation. It is worth noting that, the measured $M_s$ become smaller than the theoretical results for the Ti6 and Ti8 alloys. This unusual rapid decreasing behavior of magnetic moments of the annealed bulk samples can be attributed to multiple aspects, as shown in Fig. 3. Firstly, the original existence of Fe atoms ensures the ferromagnetic exchange interaction between Cr - Cr atoms, and its magnetic moment aligns parallelly to that of Fe atoms, contributing to larger $M_s$. Secondly, the non-magnetic (NM) FeTi secondary phase is formed by the introduced Ti and partial Fe atoms of the system, so that the magnetic moment of these Fe atoms does not contribute to $M_s$. Simultaneously, the absence of Fe atoms results in the antiparallelly aligned magnetic moment of Cr atoms, and there exists anti-ferromagnetic exchange interaction between Cr - Cr atoms, thus the magnetic moment is partly counteracted. Consequently, a remarkable decrease of the total moment is observed for the Ti6 and Ti8 alloys. To verify, the $M_s$ of ribbon samples are larger than that of the annealed bulk samples because the secondary phase is suppressed by melt spinning.

The steps of the thermogravimetry curves reflect $T_C$ in Fig. 2(c). During the introduction of Ti, $M_s$ decreases monotonically. However, $T_C$ first rises and then saturates, with a kink at Ti 4. This demonstrates that $M_s$ and $T_C$ can be remarkably tuned by Ti substitution, due to the alterations of the spin polarizations and exchange interactions. Figure 2(d) shows the $M_s$ of the Cr- and Ti- substituted $Fe_2CrV$ alloys at 5 K. The higher content of Ti and Cr atoms will result in the smaller $M_s$.

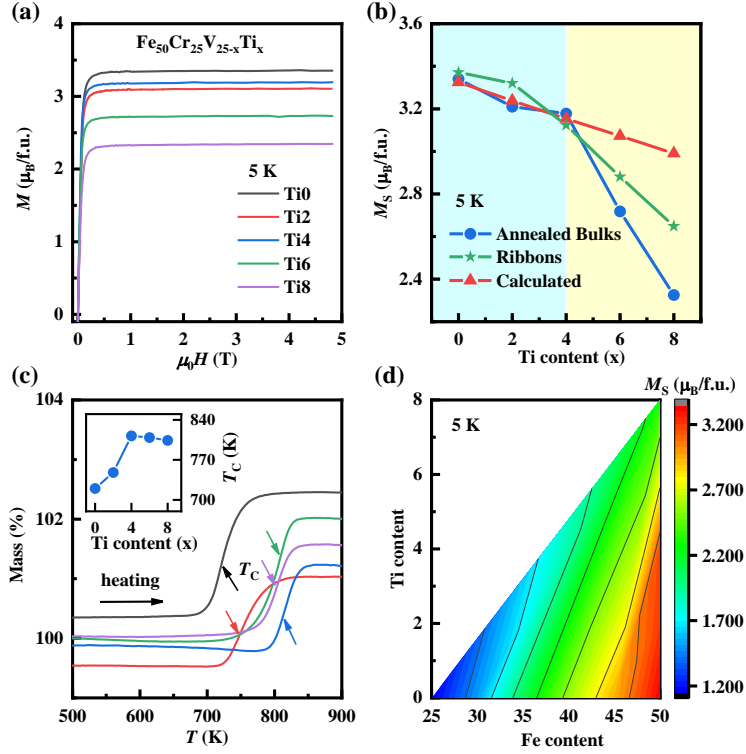

**Fig. 2.** (**a**) Magnetization curves $M(H)$ for $Fe_{50}Cr_{25}V_{25-x}Ti_x$ (x=0-8) at 5 K. (**b**) The molecular magnetic moment ($M_s$) of experimental and calculated results for different Ti content. (**c**) The TG measurements (heating) of annealed bulks. The insert shows Ti content dependence of Curie temperature ($T_C$). (**d**) the $M_s$ of different Ti and Fe content at 5 K.

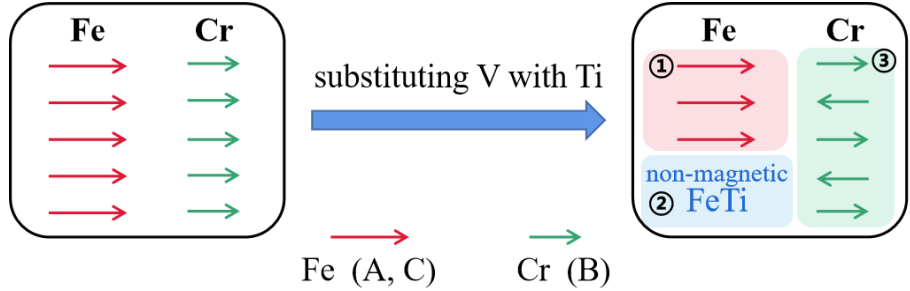

**Fig. 3.** Schematic configurations of the magnetic structure evolution for substituting V with Ti.

### 3.3. Electronic Structure

The band structures of $Fe_{50}Cr_{25}V_{25-x}Ti_x$ (x=0-8) alloys are depicted in Fig. 4(a) and Fig. 4(b). It is evident that the bands in the spin-up and spin-down channels overlap differently with $E_F$. This difference in overlap is further reflected in the density of states (DOS) at $E_F$, and the calculated total DOS is shown in Fig. 4(c). There exist some relatively flat bands in the spin-up channels along the high symmetry paths W-L and W-K, which correspond to a peak on the DOS at $E_F$. With increasing Ti doping, the flat bands shift up from the Fermi level. In contrast, fewer bands intersect the $E_F$ in spin-down channels, which corresponds to the small and finite DOS and a pseudo gap. The pronounced difference indicates the high spin polarization and a typical half-metallic characterization, which is consistent with the calculations of other all-d-metal compounds[26, 30].

It is well-established that the electronic structures and magnetic properties are mainly determined by the electrons near $E_F$. With increasing Ti doping, the peak in the majority DOS shifts to higher energy and the valley in the minority DOS moves to lower energy, which results in a decrease difference between spin-up and spin-down DOS. Therefore, the magnetic moment is decreased, which is also consistent with our results in Fig. 2(b) and Slater-Pauling rule[31]. In addition, the structure instability is also related to DOS. Based on Jahn–Teller effect[32, 33], a high DOS around $E_F$ reveals an electronic instability. For Ti8 sample, the $E_F$ locates in a peak region of DOS, thus it is more likely to induce the secondary phase.

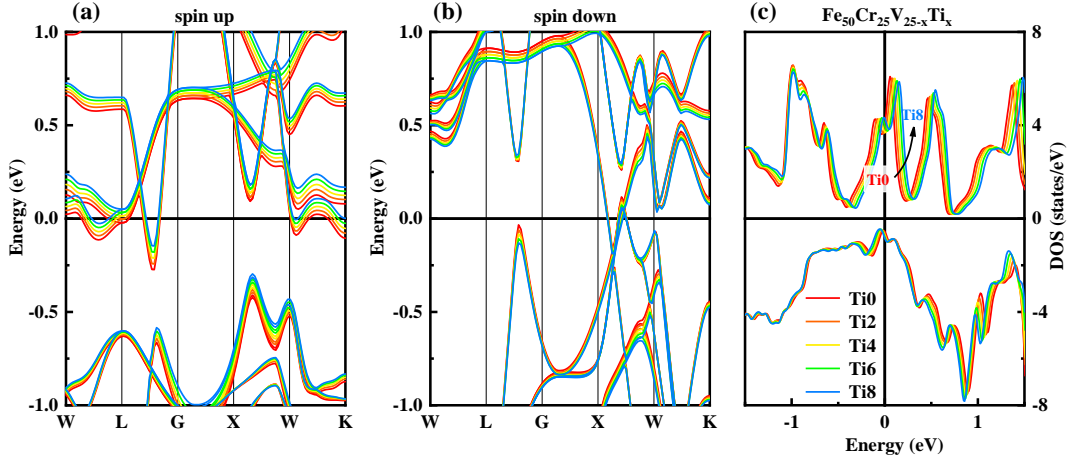

**Fig. 4.** Calculated energy dispersion of electronic bands along high symmetry paths for (**a**) spin up and (**b**) spin down. (**c**) Calculated spin-projected total DOS plots for $Fe_{50}Cr_{25}V_{25-x}Ti_x$ (x=0-8).

### 3.4. Transport Properties

We further probed the intrinsic electronic behaviors of the $Fe_2CrV$ during the substitution. Figure 5(a) shows the temperature dependence of longitudinal resistivity ($\rho_{xx}$) of $Fe_{50}Cr_{25}V_{25-x}Ti_x$ (x=0-8) samples from 300 to 5 K. The curves of $\rho_{xx}(T)$ in a zero magnetic field (0 T) are slightly larger than in 5 T, indicating the magnetoresistance is infinitely small and negative. In addition, $\rho_{xx}(T)$ exhibits a metallic-like resistivity $\left(\frac{\partial \rho_{xx}}{\partial T} > 0\right)$ throughout the measured temperature range. Generally, the electrical resistivity of a conducting material can be expressed as $\rho_{xx}(T) = \rho_0 + \rho_1(T)$. $\rho_0$ is the residual resistivity arising from the scattering of conduction electrons by lattice defects and impurities[34]. $\rho_1(T)$ is related to other bulk-scattering mechanisms. The scattering mechanisms at low temperature region are various, such as electron-electron ($\rho_{el-el}$) scattering, electron-phonon ($\rho_{el-ph}$) scattering, electron-magnon ($\rho_{el-mag}$) scattering and two-magnon ($\rho_{mag}$) scattering[35, 36]. It is noteworthy that the $T^2$-dependent $\rho_{el-mag}$ scattering is absent in the half-metallic state[37]. Two-magnon scattering processes have been considered extensively in both the broad-band case[38] (weak s-d exchange interaction) and the narrow-band case[39] (double exchange model). The resulting temperature dependences of resistivity follow $T^{\frac{7}{2}}$ and $T^{\frac{9}{2}}$, respectively. In the half-metallic ferromagnets Co$_2$FeSi[35, 36], low-temperature resistivity exhibits a power-law behavior, $\rho_{xx}(T) = T^n$ (7/2<n<9/2), in the range from 30 to 60 K. The $\rho_{xx}(T)$ data for Fe$_2$CrV are fitted with the power-law, and the fitting parameter n is 2.55, which does not fall within 7/2 and 9/2. This discrepancy may be attributed to Fe$_2$CrV not being a complete half metal, as its spin polarization is not 100%.

The $\rho_{xx}(T)$ data for all our samples fit well with the expression $\rho_{xx}(T) = \rho_0 + \rho_{el-el}T^2 + \rho_{el-ph}T^5$ below 70 K. The contribution of $\rho_{el-ph}$ scattering follows $T^5$ behavior at low-T. The variations of the fitting parameters with Ti doping are depicted in Table 2. Therefore, the resistivity is dominated by $\rho_{el-el}$ scattering and shows the behavior of Fermi liquid at low temperature region. However, it is obvious that the resistivity shows a T-linear behavior above 70 K, which indicates the resistivity is dominated by electron-phonon scattering contribution at high temperature region[40]. The similar T-linear behavior is also reported in Heusler alloys Co$_2$MnSi with p-d covalent hybridization[41]. The $\rho_{xx}(T)$ data for all our samples fit well with the expression $\rho_{xx}(T) = \rho'_0 + \rho'_{el-ph} T$ above 70 K. $\rho'_0$ is the fitting constant that depends on different Ti content. The value of $\rho_0$ increases linearly with the doping content x, while $\rho'_{el-ph}$ does not vary significantly with x. This indicates that the impurity concentration continues to increase and the intensity of electron-phonon scattering is almost constant.

| Nominal x | $\rho_0$ (μΩ cm) | $\rho_{el-el}$ ($10^{-4}$ μΩ cm K$^{-2}$) | $\rho_{el-ph}$ ($10^{-10}$ μΩ cm K$^{-5}$) | $\rho'_{el-ph}$ (μΩ cm K$^{-1}$) |
|---|---|---|---|---|
| 0 | 33.7 | 3.29 | 3.83 | 0.112 |
| 2 | 42.7 | 4.13 | 3.54 | 0.114 |
| 4 | 49.5 | 5.32 | 2.66 | 0.111 |
| 6 | 57.4 | 6.13 | 1.11 | 0.105 |
| 8 | 65.4 | 7.66 | -1.33 | 0.101 |

**Table 2** The residual resistivity ($\rho_0$), low-$T$ fitting parameters electron-electron scattering ($\rho_{el-el}$) and electron-phonon scattering ($\rho_{el-ph}$), high-$T$ fitting parameters electron-phonon scattering ($\rho'_{el-ph}$) for different Ti content of Fe$_{50}$Cr$_{25}$V$_{25-x}$Ti$_x$ system.

To obtain more information about the transport properties in this system, we performed Hall measurements. Figure 5(b) represents the magnetic field ($H$) dependence of Hall resistivity ($\rho_{yx}$) of Fe$_{50}$Cr$_{25}$V$_{25-x}$Ti$_x$. The $\rho_{yx}(H)$ exhibits the characteristic of anomalous Hall effect (AHE) and shows a linear dependence on $H$ in the high field. We considered the one-carrier model[42] for the system and performed linear fitting of the $\rho_{yx}$ above 2 T to get the slope of $\rho_{yx}(H)$ curves. The positive slope indicates that the transport properties of this system are largely dominated by hole carriers. It is a well-established principle in transport mechanism that an electron can only scatter into an unoccupied state. The probability of deep-energy-level electrons scattering into states significantly above the $E_F$ is markedly reduced. Consequently, only the electrons in the vicinity of the $E_F$ can determine the transport properties. As illustrated in Fig. 4(c), the states above $E_F$ in Ti8 are less than in Ti0 sample within a few $k_BT$s. The Ti8 sample exhibits lower unoccupied states near the $E_F$, resulting in a decrease in carrier density (n) as shown in Table 3, despite an increase in DOS at $E_F$ during the substitution. The similar phenomenon also appears in $Ni_{50-x}Fe_xMn_{35}Ti_{15}$[22]. Furthermore, the mobility (μ) at 5 K exhibit negligible variation with respect to the Ti content, maintaining approximate values of 10 $cm^2V^{-1}s^{-1}$.

Figure 5(c) represents $\rho_{yx}(H)$ curves of $Cr_{75-x}Fe_xV_{25}$ (x=25-50, denoted as Fex). The Hall conductivity is calculated by the expression, $\sigma_{xy} = \frac{\rho_{yx}}{\rho_{yx}^2 + \rho_{xx}^2}$. The AHC can be extracted from a linear extrapolation of the high-field Hall conductivity curve, see Fig. 5(b) and Fig. 5(c) inset. Figure 5(d) shows AHC for different Ti and Cr content at 5 and 300 K. The maximum value occurs at $Fe_2CrV$.

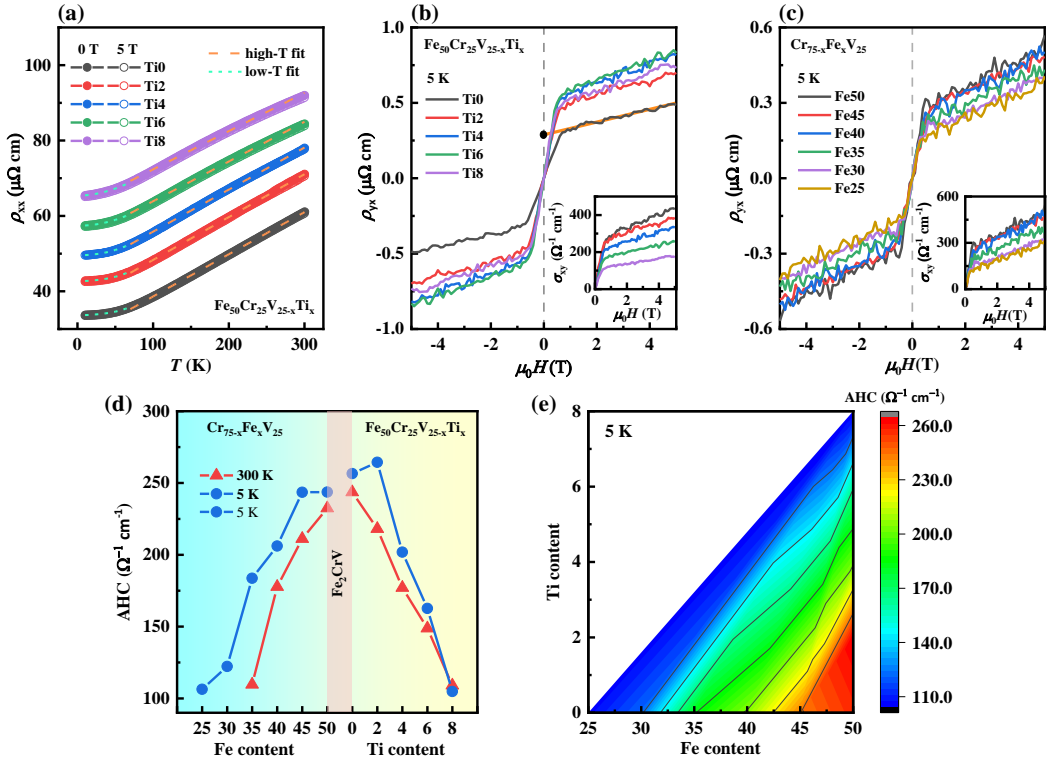

**Fig. 5.** (a) Temperature dependence of longitudinal resistivity ($\rho_{xx}$) of $Fe_{50}Cr_{25}V_{25-x}Ti_x$ (x=0-8) samples in 0 T and 5 T. The orange dotted line is the fitting curve of high temperature, using the expression as $\rho_{xx}(T) = \rho_0' + \rho_{el-ph}'T$. The light blue dotted line is the fitting curve of low temperature, using the expression as $\rho_{xx}(T) = \rho_0 + \rho_{el-el}T^2 + \rho_{el-ph}T^5$. (b), (c) Magnetic field $H$ dependence of Hall resistivity $\rho_{yx}$ at 5 K for $Fe_{50}Cr_{25}V_{25-x}Ti_x$ (denoted as Tix) and $Cr_{75-x}Fe_xV_{25}$ (denoted as Fex), respectively. The inset shows the field dependence of hall conductivity. (d) AHC of the Cr- and Ti-substituted $Fe_2CrV$ alloys at 5 K. (e) AHC for different Ti and Cr content at 5 and 300 K.

In general, AHE consists of two different microscopic mechanisms, that is, the intrinsic and extrinsic contributions[43]. The former originates from Berry curvature based on the band structure[44-46], the latter roots in the scattering process including skew scattering and side-jump effects[47, 48]. The AHE of $Fe_{50}Cr_{25}V_{25-x}Ti_x$ (x=0-8) alloys is primarily attributed to the intrinsic mechanisms. Here, we present the understanding from two different perspectives. Firstly, according to the unified model[43, 49] for $\sigma_{xx0}$ (longitudinal conductivity in 0 T) dependence of AHC, the AHE is dominated by the intrinsic mechanisms when $\sigma_{xx0}$ is between $10^4$-$10^6$ $(\Omega\, cm)^{-1}$. The $\sigma_{xx0}$ of Ti-substituted $Fe_2CrV$ alloys are in this range, as shown in Table 3, indicating that the intrinsic contributions dominate the transport. Secondly, based on the TYJ scaling[50], the intrinsic AHC remains robust when $M$ is almost unchanged. It is understood that the extrinsic AHC diminishes as temperature increases. For $Fe_{50}Cr_{25}V_{25-x}Ti_x$ system, the $T_C$ are significantly higher than room temperature. Thus, the extrinsic AHC has almost entirely decayed at 300 K, while the intrinsic AHC remains nearly unchanged, signifying the dominant contribution of intrinsic mechanisms. These two understandings should have same origin.

| Nominal x | $\sigma_{xx0}$ ($10^4\ \Omega^{-1}cm^{-1}$) | n ($10^{22}\ cm^{-3}$) | μ ($cm^2V^{-1}s^{-1}$) |
|---|---|---|---|
| 0 | 2.96 | 1.53 | 12.11 |
| 2 | 2.34 | 1.49 | 9.84 |
| 4 | 2.01 | 0.97 | 13.00 |
| 6 | 1.74 | 1.02 | 10.66 |
| 8 | 1.53 | 0.98 | 9.76 |

**Table 3** Longitudinal conductivity at 0 T ($\sigma_{xx0}$), the carrier density (n) and mobility (μ) at 5 K for different Ti content of $Fe_{50}Cr_{25}V_{25-x}Ti_x$ system.

For $Fe_{35}Cr_{40}V_{25}$ sample, the AHC at 300 K is nearly half of it at 5 K. This is because its $T_C$ (415 K) is not significantly higher than RT, leading to a decrease in both its intrinsic and extrinsic contributions. For both samples $Fe_{25}Cr_{50}V_{25}$ and $Fe_{30}Cr_{45}V_{25}$, $T_C$ (170 K and 330 K, respectively) falls below or is proximate to RT. Consequently, they exhibit only the normal Hall effect at 300 K. Figure 5(e) illustrates the AHC of the Cr- and Ti- substituted $Fe_2CrV$ alloys at 5 K. An increased content of Ti and Cr atoms results in a smaller AHC.

Next, we focus on the thermoelectric transport in our system, including Seebeck effect and anomalous Nernst effect. We selected some samples to measure the thermal conductivity (κ), Seebeck coefficient at 0 T ($S_{xx0}$), the dimensionless figure of merit (ZT) and ANC. The results obtained at 300 K are listed in Table 4. It is evident that $Fe_2CrV$ exhibits the maximum values compared to the other two samples ($Cr_2FeV$ and $Fe_{50}Cr_{25}V_{19}Ti_6$). Figure 6 depicts the $H$ dependence of Nernst thermoelectric conductivity ($\alpha_{xy}$) and Seebeck coefficient ($S_{xx}$) for $Fe_2CrV$ and $Fe_{50}Cr_{25}V_{19}Ti_6$. The magnitude of $S_{xx}$ exhibits minimal variation in response to changes in $H$, suggesting a negligible magneto-Seebeck effect (defined as $\frac{S_{xx}(B)-S_{xx}(0)}{S_{xx}(0)}$). Notably, $S_{xx}(H)$ demonstrates different behaviors at lower $H$ values (below 0.5 T). This phenomenon could potentially be attributed to the dynamics of magnetic domain walls. After saturation of magnetization, $S_{xx}(H)$ increases in a consistent slope. Notably, the saturation field obtained by the transport measurement is larger than the results of the magnetic measurement, which is attributed to the presence of a demagnetizing field when the magnetic field is applied perpendicularly to the long axis of samples in transport measurements.

| content | $\kappa$ (W/mK) | $S_{xx0}$ (μV/K) | $Z_{xx}T$ | $\alpha_{xy}^A$ (A/mK) |
|---|---|---|---|---|
| $Cr_2FeV$ | 23.9 | 9.8 | $1.9 \times 10^{-3}$ | - |
| $Fe_2CrV$ | 24.3 | 15.1 | $4.6 \times 10^{-3}$ | 0.18 |
| $Fe_{50}Cr_{25}V_{19}Ti_6$ | 16.3 | 11.8 | $3.1 \times 10^{-3}$ | 0.13 |

**Table 4** Thermal conductivity ($\kappa$), Seebeck coefficient at 0 T ($S_{xx0}$), the dimensionless figure of merit (ZT) and ANC ($\alpha_{xy}^A$) at 300 K.

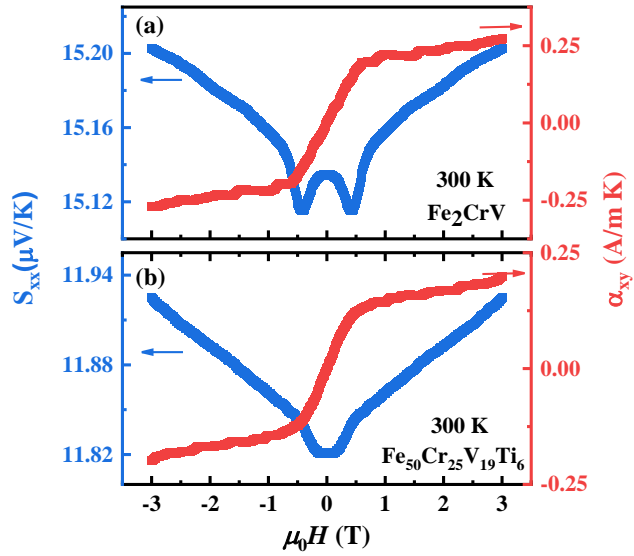

**Fig. 6.** $H$ dependence of Nernst thermoelectric conductivity ($\alpha_{xy}$) and Seebeck coefficient ($S_{xx}$) for (**a**) $Fe_2CrV$ and (**b**) $Fe_{50}Cr_{25}V_{19}Ti_6$ at 300 K.

## 4. Conclusion

The structural, magnetic and transport properties of Cr- and Ti-substituted $Fe_2CrV$ all-d-metal Heusler alloys have been investigated based on theoretical calculations and experiments. These alloys crystallize in $Hg_2CuTi$-type structure. The experimental saturation magnetization is closely consistent with theoretical calculations and Slater-Pauling rule when x < 4 for $Fe_{50}Cr_{25}V_{25-x}Ti_x$, suggesting the highly ordered atomic occupation. However, an unusual rapid decrease of saturation magnetization when x > 6 can be attributed to multiple aspects including decreased contribution of Fe atoms, non-magnetic FeTi secondary phase and anti-ferromagnetic exchange interaction between Cr - Cr atoms. In addition, the $M_s$ and $T_C$ can be remarkably tuned by substitution, based on the alterations of the spin polarizations and exchange interactions. With increasing Ti doping, the scattering of electrons by lattice defects and impurities increases linearly, while the electron-phonon scattering does not vary significantly. We also found that the AHE is dominated by the intrinsic mechanisms in Ti-substituted $Fe_2CrV$ alloys. Moreover, the maximum values of electrical and thermal transport occur at the stoichiometric $Fe_2CrV$. These behaviors provide the necessary physical understanding and application exploration for all-d-metal Heusler alloys.

**Acknowledgements**

This work was supported by the Fundamental Science Center of the National Natural Science Foundation of China (No. 52088101), the State Key Development Program for Basic Research of China (Nos. 2022YFA1403800, and 2022YFA1403400), the National Natural Science Foundation of China (No 12174426), the Strategic Priority Research Program (B) of the Chinese Academy of Sciences (CAS) (No. XDB33000000), the Synergetic Extreme Condition User Facility (SECUF), and the Scientific Instrument Developing Project of CAS (No. ZDKYYQ20210003).